\documentstyle[12pt]{article}
\begin{document}
\hoffset=-1truecm
\voffset=-2truecm
\baselineskip=18pt plus 2pt minus 2pt
\hbadness=500
\tolerance=5000
\title{\bf Comment on ``Exact Spectral Functions of a Non Fermi Liquid
in 1 Dimension''}
\author{A. A. Zvyagin \\
B.I. Verkin Institute for Low Temperature Physics and Engineering, \\
Ukrainian National Academy of Sciences, 47 Lenin Avenue, \\
Kharkov, 61164, Ukraine \\}
\date{\today}
\maketitle
\begin{abstract}
We point out the incorrect statement of the recent manuscript by K.~Penc
and B.~S.~Shastry. 
\end{abstract}

In a recent manuscript \cite{cond-mat} K.~Penc and B.~S.~Shastry wrote
in the first paragraph: 

``Schulz and Shastry \cite{SS} have introduced a new class of
gauge-coupled one-dimensional (1D) Fermi systems that are non Fermi
liquid in the sense that the momentum distribution function has a cusp
at the Fermi momentum $k_F$ rather than a jump as in a Fermi
liquid \cite{FL}. This behavior is of the sort first found by
Luttinger in the context of his study of a one-dimensional model that
is popularly known as the Luttinger model \cite{TL}. The model
introduced by Schulz and Shastry (SS) is in fact intimately connected
to the Luttinger model, and is best viewed as a reinterpretation of
Luttinger's original model as a gauge theory. Particles of different
species exert a mutual gauge potential on each other, and this is
sufficient to destroy the Fermi liquid. This model has the added
property that the charge and spin correlations are unaffected by the
interaction, owing to the `gauge' nature of interaction.''

I would like to point out that the main statement of this paragraph is
incorrect. In fact, this can be seen from the second reference from 
\cite{SS} (which is the Reply to my previous Comment \cite{com}) and
that Comment itself. It turns out that {\em all the properties,
mentioned in the cited paragraph}, namely ---  
\begin{itemize}
\item The gauge coupling between 1D Fermi systems, which produces the
non Fermi liquid behavior (the Luttinger liquid behavior); 
\item Particles of different species in that class exert a mutual
gauge potential on each other (which is sufficient to destroy the
Fermi liquid); 
\item The charge and spin correlations are unaffected by the
interaction, owing to the ``gauge'' nature of interaction   
\end{itemize} 
--- had been already known for the class of models, introduced by us
in Refs.~\cite{Zv1,Zv2} in 1992. This is why the priority of introducing
the models with these properties belongs to us. One can check that 
particles in \cite{Zv1,Zv2} of different species are connected with each
other via `gauge' potentials (reminiscent of the Peierls phase factors),
similar to the `gauge' potentials in \cite{SS}, and that we
emphasized on the Luttinger liquid behavior of our 1D models with 
this `gauge' couplings already in 1992, much earlier than \cite{SS}. 
Even the title of \cite{Zv2} is ``Exactly solvable models of an
effectively two-dimensional Luttinger liquid''. Notice that in
\cite{Zv1,Zv2} we interpreted the additional index, which distinguishes
species of particles, as a number, which enumerates 1D chains, coupled
with each other via gauge potentials. By the way, SS in their first
paper of \cite{SS} never used this interpretation, but now, in 
\cite{cond-mat}, the authors already imply that the class of models, 
introduced by SS described coupled 1D chains. 

According to the Reply \cite{SS}, SS introduced some class of 
models, which had {\em all mentioned above properties} of the class,
introduced in \cite{Zv1,Zv2}, but with the {\em additional
constraint}: Some recursion relations for `gauge' potentials have to
be satisfied. SS especially pointed out that difference, see, please,
the footnote [5] of the Reply \cite{SS}, where they wrote: ``We use the
term `class of models' in the specific sense that the members share
a common method of solution, rather than a vague sense in which many
models share certain physical properties.'' Obviously, only the properties, 
mentioned in the first paragraph of \cite{cond-mat}, cannot properly
define the class of models, introduced by SS, because they belong to
both classes: \cite{Zv1,Zv2} and \cite{SS}.   

Actually, there are two alternatives: 
\begin{itemize} 
\item (1) Either SS introduced a {\em subclass} (with some specific, 
additional properties) of the class of models \cite{Zv1,Zv2} (which had
been earlier introduced in our papers) with the {\em common}
properties, mentioned in the disputed paragraph. Certainly, one cannot
introduce any new class of models in 1998 with the same properties as
the models, introduced in 1992. 
\item (2) Or SS (cf. their Reply \cite{SS}) introduced some new class
of models, different from ours. However in this case the authors of 
the manuscript \cite{cond-mat} could clearly define the properties,
which determine only the models, introduced by SS, but which are 
{\em not} present in the class, introduced earlier in
\cite{Zv1,Zv2}. At least they could carefully distinguish between
those two classes, and not to emphasize on the properties, which
belong to the other class of models, introduced in \cite{Zv1,Zv2} 
in the definition of the models, introduced by SS. 
\end{itemize} 
In both cases the statements of the disputed paragraph \cite{cond-mat}
are wrong. I do not imply that the model, studied in \cite{cond-mat}
belongs to the class of models, introduced in \cite{Zv1,Zv2} in the
sense (2). However, clearly, when writing about the properties of the
models, introduced in \cite{SS}, the authors of \cite{cond-mat} could 
properly write about the features, which pertain {\em only to the class
introduced by SS}, but not about the ones, which had been known for
the other class of models, introduced earlier by us in
\cite{Zv1,Zv2}. It turns out that according to the definition of the
Reply \cite{SS}, all mentioned in the disputed paragraph of
\cite{cond-mat} properties have namely ``a vague sense in which many models
share certain physical properties''. 

This is why, immediately after the manuscript \cite{cond-mat} appeared
in the ArXive, I asked the authors of \cite{cond-mat} to correct the
statement of the disputed paragraph. However the authors did not agree,
i.e., they insist that the class of models with the properties,
mentioned in the cited paragraph of their work, was introduced in
\cite{SS}. But then a contradiction exists: If the statements of
the Reply \cite{SS} are correct, then the statements of the first
paragraph of the manuscript \cite{cond-mat} are obviously wrong. 

It is the goal of this my Comment to point out this contradiction.

\vfill
\eject
\end{document}